\documentclass[prd,nofootinbib,showpacs,superscriptaddress, preprint]{revtex4-1}
\usepackage[T1]{fontenc}
\usepackage{amsmath,amssymb}
\usepackage{epsfig}
\usepackage{graphicx}
\usepackage[usenames,dvipsnames]{color}
\usepackage{slashed}
\usepackage[colorlinks,citecolor=blue]{hyperref}
\usepackage{color}
\usepackage{subfig}
\begin{document}
\title{keV Neutrino Dark Matter in a Fast Expanding Universe}

\author{Anirban Biswas}
\email{anirban.biswas.sinp@gmail.com}
\affiliation{School of Physical Sciences, Indian Association for the
Cultivation of Science, 2A \& 2B Raja S.C. Mullick Road, Kolkata 700032, India}
\author{Debasish Borah}
\email{dborah@iitg.ac.in}
\affiliation{Department of Physics, Indian Institute of Technology
Guwahati, Assam 781039, India}
\author{Dibyendu Nanda}
\email{dibyendu.nanda@iitg.ac.in}
\affiliation{Department of Physics, Indian Institute of Technology
Guwahati, Assam 781039, India}

\begin{abstract}
We study the possibility of keV neutrino dark matter in the minimal $U(1)_{B-L}$ gauge extension of the standard model where three right handed neutrinos are automatically included due to the requirement of anomaly cancellations. Without considering extra additional particles or symmetries, we consider the lightest right handed neutrino to be in the keV mass range which is kinematically long lived. Due to gauge interactions, such a keV neutrino can be thermally produced in the early Universe followed by decoupling while being relativistic. The final relic abundance of such keV neutrino typically overclose the Universe requiring additional mechanism to bring it under observed limits. We propose a non-standard cosmological history where a scalar field $\phi$, that redshifts faster than radiation dominates the Universe prior to the radiation dominated era. We show that such a non-standard phase can keep the abundance of thermally generated keV neutrino dark matter within observed relic abundance. We constrain the non-standard phase, $U(1)_{B-L}$ parameters from these requirements and also briefly comment upon the observational aspects of such keV neutrino dark matter.
\end{abstract}

\maketitle
\section{Introduction} 
\label{sec:intro}
The energy budget of our present Universe is constituted, in significant amount, by a non-luminous, non-baryonic form of matter, known as dark matter (DM). According to the latest data from the Planck satellite mission \cite{Ade:2015xua}, DM accounts for around $26\%$ of our Universe at present, which is often quoted in terms of density parameter $\Omega$ and $h = \text{(Hubble Parameter)}/(100 \;\text{km} \text{s}^{-1} \text{Mpc}^{-1})$ as
\cite{Ade:2015xua}
\begin{equation}
\Omega_{\text{DM}} h^2 = 0.1186\pm0.0020.
\label{dm_relic}
\end{equation}
Apart from such cosmology experiments, there are evidence from astrophysics based experiments as well, like the galaxy cluster observations by Fritz Zwicky \cite{Zwicky:1933gu} back in 1933, observations of galaxy rotation curves in 1970's \cite{Rubin:1970zza}, the more recent observation of the bullet cluster \cite{Clowe:2006eq} among others. While the particle origin of DM is not yet known, there have been a wide range of possibilities discussed so far, all within beyond standard model (BSM) frameworks. Although among the standard model (SM) particles, neutrinos satisfy some of the criteria for being a good DM candidate, yet they remain relativistic at the epoch of freeze-out as well as matter radiation equality, giving rise to hot dark matter (HDM) which is ruled out by both astrophysics and cosmology observations. Among the BSM proposals, the weakly interacting massive particle (WIMP) paradigm is the most popular one. In this framework, a dark matter candidate typically having mass in the GeV-TeV scale and interaction rate similar to electroweak interactions can give rise to the correct dark matter relic abundance, a remarkable coincidence often referred to as the \textit{WIMP Miracle}. Such interactions enable the WIMP DM to be produced in thermal equilibrium in the early Universe and eventually its number density gets frozen out when the rate of expansion of the Universe takes over the interaction rates. Such DM candidates typically remain non-relativistic at the epoch of freeze-out as well as matter radiation equality and belong to the category of Cold Dark Matter (CDM). 

Since WIMP dark matter has sizeable interactions with the SM particles, one also expects that the same interactions will enhance the testability of such DM candidates as they can scatter off nuclei kept in a detector. Several direct detection experiments have been designed with this spirit but till date, no such DM-nucleon scattering has been observed in any of the experiments. The most recent dark matter direct detection experiments like LUX, PandaX-II and Xenon1T have also reported their null results \cite{Akerib:2016vxi, Tan:2016zwf, panda2017, Aprile:2017iyp}. Similar null results have been reported by DM searches at collider experiments as well, for example a review of DM searches at the large hadron collider (LHC) can be found in \cite{Kahlhoefer:2017dnp}. Such null results for WIMP DM have motivated the particle physics community to look for other alternatives. One interesting possibility is the warm dark matter (WDM) scenario where the DM remains mildly relativistic at the epochs of matter radiation equality and hence keeping such DM candidates at intermediate stage between HDM and CDM. Such DM candidates typically have masses in the keV regime which is intermediate between su-eV scale masses of HDM and GeV-TeV scale masses of CDM. To be more appropriate, this classification is primarily done on the basis of free streaming lengths (FSL), the distance through which a DM particle can freely propagate. For detailed calculation of free streaming lengths, please refer to \cite{Boyarsky:2008xj, Merle:2013wta}. A popular WDM candidate is a right handed neutrino, singlet under the SM gauge symmetry (and hence called sterile), having tiny mixing with the SM neutrinos leading to a long lifetime. For a recent review on such keV sterile neutrino DM, please see \cite{Adhikari:2016bei}. WDM can be motivating not only from the null results at the WIMP or CDM sector, but it can also solve some astrophysical structure related problems in CDM paradigm. For a recent review on such astrophysical problems related to small scale structure, please refer to \cite{Bullock:2017xww}. Due to the difference in their FSL, these different categories of DM can give rise to different structure formation which can be tested at galaxy survey experiments. For example, due to small FSL, the structures in a CDM dominated Universe keep forming till scales as low as the solar system which is in disagreement with observations at small scales  \cite{Bullock:2017xww}. HDM, on the other hand, erases all small scale structure due to its large free streaming length, disfavouring the bottom up approach of structure formation. WDM can therefore act as a balance between the already ruled out HDM possibility and the CDM paradigm having issues with small scale structures. Although such WDM can not be detected at typical direct search experiments or the LHC, it can have interesting signatures at indirect search experiments. For example, a sterile neutrino WDM candidate having mass 7.1 keV can decay on cosmological scales to a photon and a SM neutrino, providing an origin to the unidentified 3.55 keV X-ray line reported by two independent analysis \cite{Bulbul:2014sua} and \cite{Boyarsky:2014jta} of the data collected by the XMM-Newton X-ray telescope.

Typically, keV WDM candidates are long lived having tiny mixing or couplings with the SM particles, making it difficult for thermal production in the early Universe. Several interesting proposals have been put forward addressing the issue of keV DM production in the early Universe, a summary of which can be found in the recent review \cite{Adhikari:2016bei}. On the other hand, such keV WDM can be thermally generated in the early Universe, if there exists a sizeable portal between DM and SM sectors. Interestingly, such a possibility exists naturally, in $U(1)_{B-L}$ extensions of the SM, where $B$ and $L$ correspond to baryon and lepton numbers respectively. This minimal and economical model generating non-zero neutrino mass has been studied for a long time \cite{Wetterich:1981bx, Mohapatra:1980qe, Marshak:1979fm, Masiero:1982fi, Mohapatra:1982xz, Buchmuller:1991ce}. The most interesting feature of this model is that the inclusion of three right handed neutrinos, as it is done in type I seesaw mechanism of generating light neutrino masses, is no longer a choice but a necessity due to the requirement of the new $U(1)_{B-L}$ gauge symmetry to be anomaly free. We note that type I seesaw mechanism \cite{Minkowski:1977sc, GellMann:1980vs, Mohapatra:1979ia, Schechter:1980gr} is the most minimal way of generating tiny neutrino masses and mixing as suggested by experimental observations \cite{Olive:2016xmw} where three right handed neutrinos, singlet under the SM gauge symmetry are included. In typical WIMP or CDM studies within $U(1)_{B-L}$ model with type I seesaw, one usually introduces additional scalar field or an additional discrete symmetry to stabilise one of the right handed neutrinos. However, if the lightest right handed neutrino has mass in the keV regime with tiny Yukawa couplings to the SM neutrinos, one can have a long lived WDM candidate in this model without any additional symmetries. Apart from explaining the origin of light neutrino masses, the model also provides a natural way for DM production in the early Universe through $U(1)_{B-L}$ portal interactions. However, for generic values of $U(1)_{B-L}$ gauge couplings and gauge boson masses near the electroweak regime, one usually finds that the lightest right handed neutrino which decouples from the rest of the plasma while being relativistic, typically gives rise to overproduction of DM. For example, this was studied within a broader class of models having $U(1)_{B-L}$ gauge symmetry by the authors of \cite{Nemevsek:2012cd, Bezrukov:2009th}. In such a scenario, the abundance of WDM can be brought to the observed DM limits by late time entropy dilution mechanism due to the late decay of heavier right handed neutrinos \cite{Scherrer:1984fd}. In another $U(1)_{B-L}$ model with inverse seesaw mechanism for light neutrino masses \cite{El-Zant:2013nta}, such an overproduced keV DM was diluted by late decay of moduli fields. In such scenarios, we need to fine tune several Yukawa couplings in order to keep the mixing of WDM with light neutrinos small as well as to allow the late decay of heavier right handed neutrinos, after the decoupling of the lightest right handed neutrino.

In this article, we propose a different way to bring the overproduced DM density to the observed limits. We assume that, prior to the era of the big bang nucleosynthesis (BBN)  that is typically around 1 s after the big bang, the Universe was dominated by some scalar field $\phi$ instead of radiation such that the energy density red-shifts with the scale factor $a$ as follows
\begin{equation}
\rho_{\phi}\propto a^{-(4+n)}
\end{equation}
where $n>0$. Such a possibility (coined as fast expanding Universe) where the energy density at early epochs redshifts faster than radiation leading to $\phi$ domination at early Universe but negligible at later epochs was first discussed in the context of WIMP dark matter by the authors of \cite{DEramo:2017gpl}. This is also extended to non-thermal or freeze-in DM models in \cite{DEramo:2017ecx}. In the above expression, $n=0$ corresponds to the usual radiation dominated Universe. Here we consider the same $\phi$ dominated phase in the early Universe and find that the keV sterile neutrino DM in $U(1)_{B-L}$ model can be thermally produced in the early Universe but at the same time can be prevented from being overproduced. We constrain the parameters in the $U(1)_{B-L}$ sector along with the parameter $n$ dictating the redshift of $\phi$ energy density from the requirements of producing the correct DM relic abundance in the present Universe.

This article is organised as follows. In section \ref{sec:model} we briefly discuss the minimal  $U(1)_{B-L}$ model followed by the discussion of keV sterile neutrino DM in such model in section \ref{sec:kevmodel}. We briefly summarise the cosmology of a fast expanding Universe dominated by $\phi$ field in section \ref{sec:fexpun} followed by our results for keV sterile neutrino DM in section \ref{sec:kevfast}. We briefly discuss some detection prospects of such keV neutrino DM in section \ref{sec:det} and finally conclude in section \ref{sec:conc}.

\section{The Minimal $U(1)_{B-L}$ Model}
\label{sec:model}
As pointed out earlier, the $B-L$ gauge extension of the SM is a very natural and minimal possibility as the corresponding charges of all the SM fields under this new symmetry is well known. Also, the SM has an accidental $U(1)_{B-L}$ global symmetry motivating one to explore the scenario where this can be uplifted to a gauge symmetry. However, a $U(1)_{B-L}$ gauge symmetry with only the SM fermions is not anomaly free. This is because the triangle anomalies for both $U(1)^3_{B-L}$ and the mixed $U(1)_{B-L}-(\text{gravity})^2$ diagrams are non-zero. These triangle anomalies for the SM fermion content turns out to be
\begin{align}
\mathcal{A}_1 \left[ U(1)^3_{B-L} \right] = \mathcal{A}^{\text{SM}}_1 \left[ U(1)^3_{B-L} \right]=-3  \nonumber \\
\mathcal{A}_2 \left[(\text{gravity})^2 \times U(1)_{B-L} \right] = \mathcal{A}^{\text{SM}}_2 \left[ (\text{gravity})^2 \times U(1)_{B-L} \right]=-3
\end{align}
Remarkably, if three right handed neutrinos are added to the model, they contribute $\mathcal{A}^{\text{New}}_1 \left[ U(1)^3_{B-L} \right] = 3, \mathcal{A}^{\text{New}}_2 \left[ (\text{gravity})^2 \times U(1)_{B-L} \right] = 3$ leading to vanishing total of triangle anomalies. This is the most natural and economical $U(1)_{B-L}$ model where the fermion sector has three right handed neutrinos $N_{Ri}$ ($i=1,2,3$) apart from the usual SM fermions \footnote{For other exotic and non-minimal solutions to such anomaly cancelation conditions, please refer to \cite{Montero:2007cd, Wang:2015saa,Patra:2016ofq, Nanda:2017bmi,Bernal:2018aon} and references therein.}. A singlet scalar has also been introduced to break the $U(1)_{B-L}$ gauge symmetry spontaneously and to provide mass to the RHNs. The particle content of this minimal model is shown in table \ref{tab1}.

\begin{table}[htbp]
\caption{Particle content}
\centering
\vspace{0.3cm}
\begin{tabular}{|l|l|l|l|l|l|}
\hline
$\psi$ &  SU(3)$_c$ & SU(2)$_L$ & U(1)$_Y$ & U(1)$_{B-L}$   \\
\hline
\hline
q$_L$ & 3 & 2 & $\frac{1}{6}$ & $\frac{1}{3}$ \\
\hline
u$_R$ & 3 & 1 & $\frac{2}{3}$ & $\frac{1}{3}$ \\
\hline
d$_R$ & 3 & 1 & -$\frac{1}{3}$ & $\frac{1}{3}$ \\
\hline
$\ell_L$ & 1 & 2 & -$\frac{1}{2}$ & -1 \\
\hline
$e_R$ & 1 & 1 & -1 & -1  \\
\hline
N$_R$ & 1 & 1 & 0 & -1 \\
\hline
\hline
H & 1 & 2 & $\frac{1}{2}$ & 0 \\
\hline
$\chi$ & 1 & 1 & 0 & 2 \\
\hline
\end{tabular}
\label{tab1}
\end{table}

The Yukawa Lagrangian can be written as 
\begin{eqnarray}\nonumber
\mathcal{L}_Y &=& \sum_{j,k=1}^{3}-y_{jk}^d \bar{q}_{jL}d_{kR} H-y_{jk}^u \bar{q}_{jL}u_{kR} \tilde{H} - y_{jk}^e \bar{\ell}_{jL}e_{kR} H - y_{jk}^\nu \bar{\ell}_{jL}N_{kR} \ \tilde{H} \\
&&- y_{jk}^M (\bar{N}_{R})_j^c \ N_{kR} \ \chi + {\rm h.c.}
\label{eqn2}
\end{eqnarray}
and the scalar potential of the model will be 
\begin{eqnarray}\nonumber
V(H,\chi) &=& -\mu_H^2 H^{\dagger}H- \mu_\chi^2 \chi^{\dagger}\chi + \lambda_H (H^{\dagger}H)^2+ \lambda_\chi (\chi^{\dagger}\chi)^2+ \lambda_{H\chi} (H^{\dagger}H)(\chi^{\dagger}\chi).
\end{eqnarray}
We choose the mass squared terms of H and $\chi$ to be negative so that the neutral components of them get nonzero vacuum expectation value (vev). We can present the scalar fields as
$$\langle H \rangle = \frac{1}{\sqrt{2}}\begin{pmatrix}0\\
v+h\end{pmatrix}, \;\; \langle \chi \rangle = \frac{u+\phi}{\sqrt{2}} $$
The minimization condition gives 
$$\mu_{H}^2=\lambda_{H}v^2+\frac{\lambda_{H\chi}}{2}u^2$$ 
$$\mu_{\chi}^2=\lambda_{\chi}u^2+\frac{\lambda_{H\chi}}{2}v^2$$ 

The neutral scalar mass matrix becomes
$$\begin{pmatrix} 2 \lambda_H  v^2 & \lambda_{H\chi} u v\\
\lambda_{H\chi} u v & 2 \lambda_\chi  u^2\end{pmatrix}$$

The mass eigenstates H$_{1}$ and H$_2$ are linear combinations of
h and $\phi$ and can be written as
\begin{eqnarray}\nonumber
H_{1}= h \cos \alpha - \phi \sin \alpha \\
H_{2}= h \sin \alpha + \phi \cos \alpha
\end{eqnarray}
Where 
\begin{eqnarray}
\tan 2\alpha=\frac{\lambda_{H\chi} u v}{\lambda_H v^2-\lambda_\chi u^2}
\end{eqnarray}

Mass terms of various scalar particles as derived from the
potential are
\begin{eqnarray}\nonumber
m_{H_1}^2= \lambda_H v^2+\lambda_\chi u^2 - \sqrt{(\lambda_H v^2-\lambda_\chi u^2)^2+(\lambda_{H\chi} u v)^2}\\ \nonumber
m_{H_2}^2= \lambda_H v^2+\lambda_\chi u^2 + \sqrt{(\lambda_H v^2-\lambda_\chi u^2)^2+(\lambda_{H\chi} u v)^2}\\
\hspace{1.6cm}
\end{eqnarray}

From the kinetic term of the scalars we can write the mass of the new gauge boson as
$$m_{Z_{BL}}=2\,g_{BL}\,u$$
Neutrino mass arises naturally through the type I seesaw mechanism. The neutral fermion mass matrix in $(\nu, N_R)$ basis can be written as
\begin{equation}
M=\begin{pmatrix}
0 & M_D \\
M^T_D & M_R
\end{pmatrix}
 \end{equation}
where $M_D = \frac{y^{\nu} v}{\sqrt{2}}$ is the Dirac neutrino and $M_R = \sqrt{2}  y^M u$ is the right handed neutrino mass matrix. Assuming $M_D \ll M_R$, the light neutrino mass matrix can be found as
\begin{equation}
M_{\nu} = -M_D M^{-1}_R M^T_D 
\end{equation}
which can give rise to the sub-eV scale neutrino mass as well as large leptonic mixing, if the strength and structure of $M_D, M_R$ are appropriately chosen.

\section{keV dark matter in $U(1)_{B-L}$ model}
\label{sec:kevmodel}

In the minimal $U(1)_{B-L}$ model discussed above, none of the BSM particles are stable as there is no remnant symmetry after spontaneous breaking of $U(1)_{B-L}$ gauge symmetry to protect it from decaying into SM particles. The only possibility, without introducing additional symmetry or particles, to have DM in this model is to consider the lightest right handed neutrino to be cosmologically long lived. This is possible if the light neutrino mass is below the electron mass threshold, thereby kinematically forbidding the tree level decay modes. If the lightest right handed neutrino has mass around a few keV and has tiny mixing with the SM neutrinos (which can be ensured by very small Yukawa couplings), it can behave like a long lived DM, falling in the WDM regime, as we discuss below. Such a DM candidate can however, decayon cosmological scales into a photon and a SM neutrino at radiative level, giving rise to interesting indirect detection signatures which we will comment upon later.
\begin{figure}[!!!!!!h]
\centering
\begin{tabular}{cc}
\epsfig{file=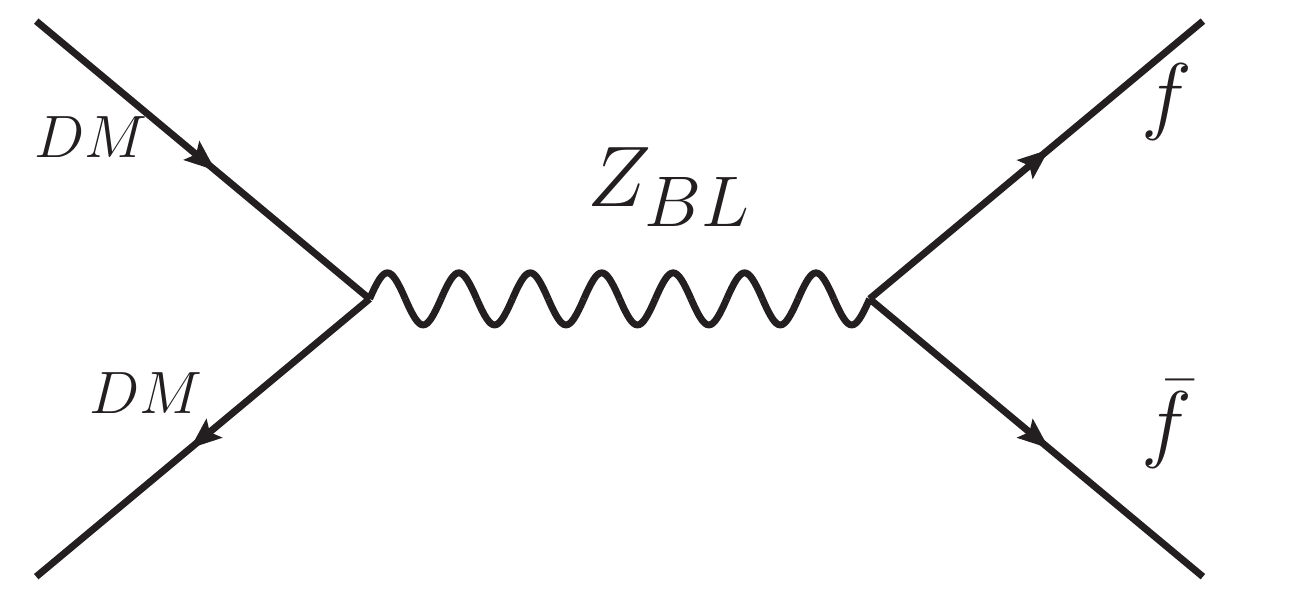,width=0.4\textwidth,clip=}
\end{tabular}
\caption{Dominant dark matter annihilation channel in this model.}
\label{fig1}
\end{figure}

\begin{figure}[!!!!!!h]
\centering
\begin{tabular}{cc}
\epsfig{file=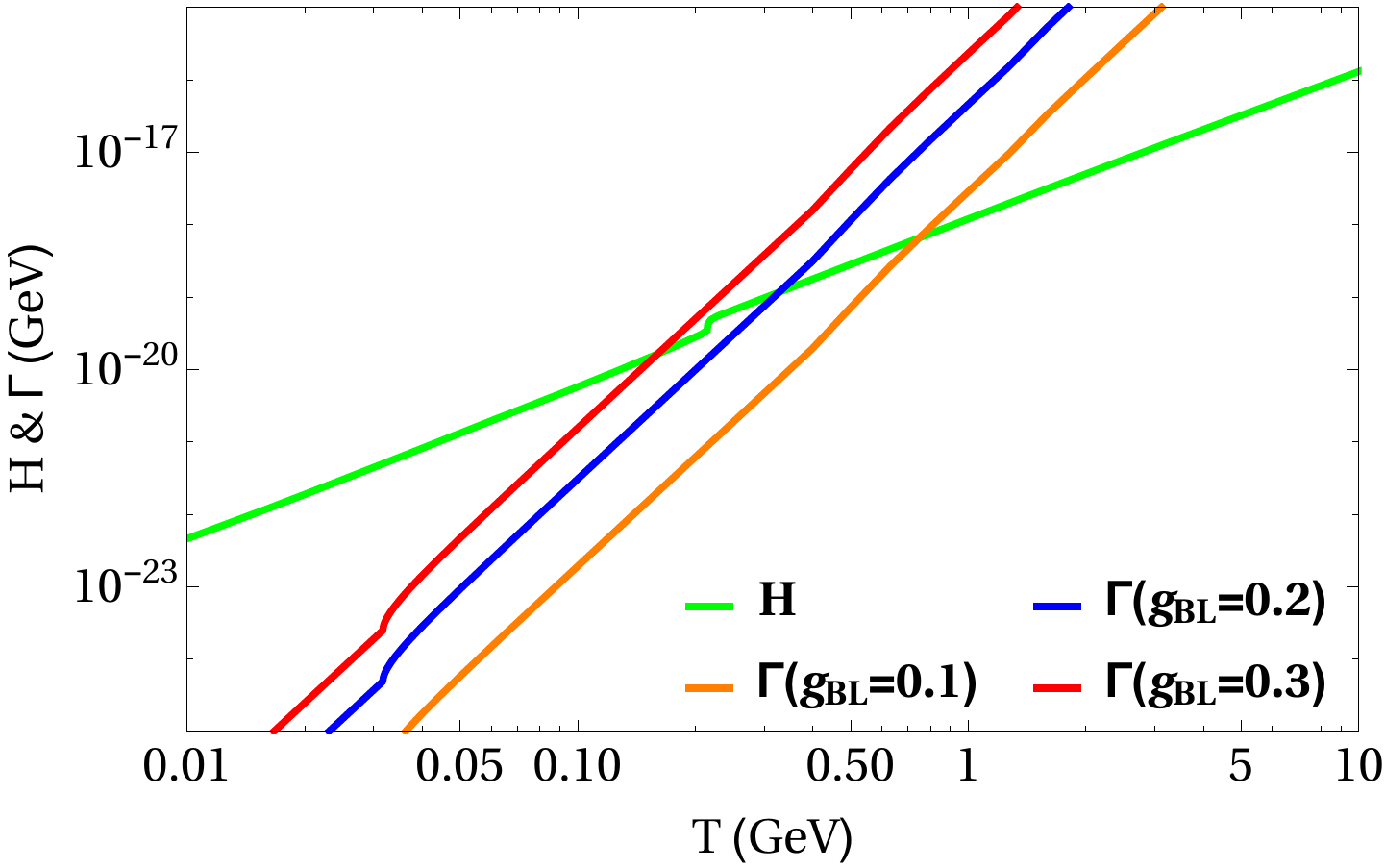,width=0.75\textwidth,clip=}
\end{tabular}
\caption{DM interaction rate versus the rate of expansion of the Universe. We have kept fixed m$_{Z_{BL}}$=3 TeV.}
\label{fig1a}
\end{figure}

We now consider the lightest right handed neutrino (denoted by $N_{1}$) mass to be 5 keV and calculate its relic abundance. Unlike in typical keV sterile neutrino DM models, here the relic abundance of $N_1$ can be calculated in a much simpler way as the $U(1)_{B-L}$ gauge interactions bring $N_1$ into thermal equilibrium in the early Universe. This happens through $U(1)_{B-L}$ portal interactions between DM and SM fermions as shown by the Feynman diagram in figure \ref{fig1}. There can be Yukawa or Higgs portal interactions as well, which we are assuming to be negligible compared to the gauge portal. This is justified as the Yukawa coupling of $N_1$ with the singlet scalar $\chi$ is suppressed as $\sim \mathcal{O}(\rm keV/\rm TeV)$. As the Universe expands, there will be an epoch at which the DM-SM interaction rate $\Gamma$ will fall below the rate of expansion of the Universe, measured in terms of the Hubble parameter $H$.
Here, the interaction rate $\Gamma$ is defined as
\begin{eqnarray}
\Gamma = n_{\rm DM} \langle {\sigma {\rm v}}\rangle \,,
\label{gamma}
\end{eqnarray}
where $n_{\rm DM}$ is the number density of $N_1$ which we have considered to be equal to the
the number density of a relativistic fermion following Fermi-Dirac distribution i.e.
\begin{eqnarray}
n_{\rm DM} = \dfrac{3}{4} \dfrac{\zeta(3)}{\pi^2} g_{\rm DM}\,T^3\,,
\label{ndm}
\end{eqnarray}
$g_{\rm DM}$ being the internal degrees of freedom of $N_1$ and $\zeta(3)= 1.20206$
is the Riemann zeta function. The expression of $\sigma {\rm v}$ for the annihilation
process shown in figure \ref{fig1} is given by
\begin{eqnarray}
\sigma {\rm v} = \dfrac{g_{BL}^4\,q_N^2\,q_f^2\,n_c}{64 \pi\,s_{E}^{3/2}}
\sqrt{s_{E}-4\,m_f^2}\dfrac{\frac{32}{3}(s_{E}-4m^2_{N_1})(s_{E}+2m^2_f)}{(s_{E}-m^2_{Z_{BL}})^2+
(\Gamma_{Z_{BL}}\,m_{Z_{BL}})^2}\,.
\label{sigmav} 
\end{eqnarray}

Where, $q_N$ and $q_f$ are the B-L charges for right handed neutrino and SM fermion $f$ respectively
while $n_c$ being the colour charge of $f$ and $\Gamma_{Z_{BL}}$ is the total decay width
of $Z_{BL}$, expression of which can be found in Ref. \cite{Biswas:2016bfo}. In the above expression $s_E$ is the Mandelstam variable which can be written as the square of energy in the centre of mass frame. As in our case, the particles are relativistic at the time of decoupling, we have replaced $s_E$ by 4$(3.151 T)^2$ where $T$ is the temperature and $3.151T$ is the average momentum of a relativistic fermion following the Fermi-Dirac distribution.  Further,
the Hubble parameter in the radiation dominated era of standard cosmological scenario
is given by
\begin{eqnarray}
H(T) = \dfrac{\pi}{3\,\sqrt{10}}\dfrac{\sqrt{g_{*}(T)}}{M_{Pl}}\,T^2\,.
\label{hub_rad}
\end{eqnarray}
Here, $g_{*}(T)$ is the relativistic degrees of freedom present at temperature $T$ while $M_{Pl}={1}/{\sqrt{8\pi\,G}}$ is
the reduced Planck mass.\,\,Using,\,\,equations\,\,\eqref{gamma}-\eqref{hub_rad},     
we compare the DM-SM interaction rate with Hubble rate in figure \ref{fig1a} for a benchmark choice of $Z_{BL}$ mass $m_{Z_{BL}}=3$ TeV and three different choices of gauge couplings $g_{BL}$. As expected, for larger values of gauge couplings, the DM can remain in equilibrium till later epochs. For the most optimistic choice of $g_{BL}$, allowed for the chosen $Z_{BL}$ mass, the DM decouples at a temperature around 400 MeV. The situation is therefore similar to the way SM neutrinos decouple in the early Universe, while remaining relativistic and very different from the WIMP freeze-out case. Of course, the exact decoupling temperature ($T_f$) will depend on the choices of model parameters such as coupling constants and the mass of the different particles etc. but for most of the generic choices of parameters, the keV sterile neutrino DM will decouple while being relativistic. The calculation of final relic abundance of such species is fairly simple and following the standard prescription given by Kolb and Turner \cite{Kolb:1990vq}, the present abundance of $N_1$ can be written as 
\begin{equation}
\Omega_{N_1}h^2=76.4 \bigg[\frac{3g}{4}\frac{1}{g_{*s}(x_{f})}\bigg]\left( \frac{M_{N_1}}{\rm keV} \right),
\label{eqn6}
\end{equation} 
where $M_{N_1}$ is the mass of $N_1$, $g$ is the internal degrees of freedom and the factor 3/4 comes from the fact that $N_1$ is a fermion. In the above equation $g_{*s}$ represents the number of relativistic entropy degrees of freedom at the epoch of $N_1$ decoupling $x_f = M_{N_1}/T_f$. The above expression says that for the species like $N_1$ which decouple while being relativistic, the present abundance depends upon the mass and $g_{*s} (x_f)$. Also at high temperatures $T> \mathcal (\rm MeV)$, we can write $g_{*s} = g_*$, the relativistic energy degrees of freedom. If $N_1$ decouples after the QCD phase transition which corresponds to $g_* \approx 10.75$, the $N_1$ abundance for $M_{N_1} = 5$ keV will be 
\begin{align}
\Omega_{N_1}h^2 &=76.4 \bigg[\frac{3\times 2}{4}\frac{1}{10.75}\bigg] \times 5 \nonumber \\
& \approx 53.302.
 \label{eqn7}
\end{align}
Which shows that if $N_1$ is DM, then it is overproduced by around 500 times compared to the DM abundance given by \eqref{dm_relic}. Even if the decoupling occurs above the electroweak symmetry breaking scale so that $g_{*s}(x_f) \approx 107$, the abundance of $N_1$ will be much more than the observed DM, overclosing the Universe. The standard procedure is to consider entropy dilution after freeze-out to bring down the abundance of $\Omega_{N_1} \leq \Omega_{\text{DM}}$. Late decay of heavier right handed neutrinos like $N_2$ can release such entropy. Such a decay should however occur before the big bang nucleosynthesis temperature $T_{\text{BBN}} \sim \mathcal{O}$(MeV) in order to be consistent with successful BBN predictions. Such late decay of long lived particles can release extra entropy and dilute the abundance of keV dark matter to bring it into the observed limit \cite{Scherrer:1984fd}. As utilised for left right symmetric models by the authors of \cite{Nemevsek:2012cd, Bezrukov:2009th}, it severely constrains the spectrum of heavier neutrino masses $M_{N_2}, M_{N_3}$ as well as their Yukawa couplings with SM neutrinos. Unlike in left right symmetric models having additional sources for light neutrino masses, in our minimal model such restrictions may be in tension with satisfying correct light neutrino masses and mixings. It should be noted that by choosing $N_1$ mass in the keV regime and having very tiny coupling with the SM neutrinos for long-livedness, we are already in a scenario where only $N_2, N_3$ effectively take part in the type I seesaw mechanism giving rise to an almost vanishing lightest neutrino mass. The details of such a calculation of entropy dilution and agreement with neutrino data within minimal $U(1)_{B-L}$ model can be found elsewhere. Instead of pursuing the entropy dilution mechanism here, we consider a different cosmological history which can also bring such thermally overproduced relic below the observed limits. This is the main topic of discussion in the upcoming sections.

\section{Fast expanding universe}
\label{sec:fexpun}
In the standard model of cosmology, the Universe prior to matter domination, was filled with radiation whose energy density scales with the size of the Universe (denoted by the scale factor $a$) as
$$ \rho_{\rm rad} (t) = \rho(t_0) \left ( \frac{a(t_0)}{a(t)} \right)^4.$$
Then there arises an epoch during BBN and recombination when matter starts to dominate over radiation. Since BBN predictions are well tested and we know for sure that the energy budget of the Universe was radiation dominated at that stage. However, there exists no such experimental evidence that suggest that the Universe to be radiation dominated throughout the epochs from the end of inflation $(T_{\rm RH})$ to BBN. Since, DM physics may have some non-trivial phase during this long era $T_{\rm BBN} < T <T_{\rm RH} $, deviation from a purely radiation dominated Universe or standard cosmological history can have non-trivial impact on final relic abundance of DM. Specially, WIMP DM freeze-out typically occurs in this era and hence non-standard cosmology can definitely affect their abundance. Deviation from standard cosmological history and its implications for WIMP type DM's thermal relic abundance have been studied for a long time, for example see \cite{McDonald:1989jd, Kamionkowski:1990ni, Chung:1998rq, Giudice:2000ex, Moroi:1999zb, Allahverdi:2002nb, Allahverdi:2002pu,  Acharya:2009zt, Davoudiasl:2015vba} and references therein. In this work, we consider a broad class of such non-standard cosmology where the Universe at very high temperatures was dominated by a field $\phi$ whose energy density varies with the scale factor $a$ as 
\begin{equation}
\rho_{\phi}\propto a^{-(4+n)}, \; n>0
\label{eqn8}
\end{equation}
where $n=0$ will restore the usual radiation dominated Universe or the standard cosmology.
As mentioned before, this was recently proposed in the context of thermal WIMP DM in \cite{DEramo:2017gpl} and later extended to non-thermal or freeze-in DM models in \cite{DEramo:2017ecx}. Here we use such non-standard cosmological history for a particular class of DM models where a keV neutrino plays the role of DM. As obvious from the above scaling of energy density, the energy density of the Universe will be dominated more and more by the $\phi$ field as we go to higher temperatures or smaller size $(a)$. However, in order to reproduce the radiation dominated Universe around the BBN era, there should be an epoch before BBN where radiation must take over from $\phi$. Adopting the notation of \cite{DEramo:2017gpl}, this epoch is identified by temperature $T_r$ where $\rho_{\phi} (T_r) = \rho_{\rm rad} (T_r)$. Thus, the non-standard cosmological phase corresponds to $T_r < T < T_{\rm RH}$ where $T_r > T_{\rm BBN}$.

Now, the total energy density of the Universe in the very early epochs can be written as
\begin{equation}
\rho(T)=\rho_{\rm rad}(T)+\rho_{\phi}(T)
\end{equation}
where the usual radiation energy density $\rho_{\rm rad}$ can be written as
\begin{equation}
\rho_{\rm rad}=\frac{\pi^2}{30}g_{*}(T)\,T^4
\label{eqn10}
\end{equation}
with $g_*(T)$ being the relativistic energy degrees of freedom given by 
$$ g_{*} (T)= \sum_{i \in \text{boson}} \left ( \frac{T_i}{T} \right)^4 g_i + \frac{7}{8} \sum_{i \in \text{fermion}} \left( \frac{T_i}{T} \right)^4 g_i. $$
In the above expression $g_i$ denotes the internal degrees of freedom for species $i$. Now we can express the $\rho_{\phi}$ as a function of temperature by using the fact that the total entropy must be conserved in a comoving volume. The total entropy $S = s a^3=$ constant, where the entropy density for radiation is
\begin{equation}
s(T)= \frac{2\pi^2}{45}g_{*s}(T)\,T^3
\label{eqn11}
\end{equation}
with $g_{*s}$ being the number of relativistic entropy degrees of freedom. Since the entropy for radiation and the $\phi$ field can be conserved independently, in the absence of any interaction between them, once can use entropy conservation for radiation to calculate the ratio of scale factor at $T=T_r$ and at any higher temperature $T>T_r$. This gives 
\begin{equation}
g_{*s}(T)T^3 a^3 (T) = g_{*s}(T_r)T^3_r a^3 (T_r)  \implies \frac{a(T_r)}{a(T)} = \left( \frac{g_{*s}(T)}{g_{*s}(T_r)} \right)^{1/3} \frac{T}{T_r}.
\label{eqn11a}
\end{equation}
By using equation \eqref{eqn8} and equation \eqref{eqn11a} one can express $\rho_{\phi}$ as function of temperature 
\begin{equation}
\rho_{\phi} (T) = \rho_{\phi}(T_r)\bigg(\frac{g_{*s}(T)}{g_{*s}(T_r)}\bigg)^{(4+n)/3} \bigg( \frac{T}{T_r}\bigg)^{(4+n)}
\end{equation}
and this leads us to the total energy density as 
\begin{equation}
\rho (T) = \rho_{\rm rad} (T)+\rho_{\phi} (T) = \rho_{\rm rad} (T)\Bigg[1 + \frac{g_{*}(T_r)}{g_{*}(T)}\bigg(\frac{g_{*s}(T)}{g_{*s}(T_r)}\bigg)^{(4+n)/3} \bigg( \frac{T}{T_r}\bigg)^{n} \Bigg]
\label{eqn13}
\end{equation}
Using equation \eqref{eqn10} in equation \eqref{eqn13} we can write 
\begin{equation}
\rho(T)=\frac{\pi^2}{30}g_{*}^{\rm eff}(T)\,T^4
\label{eqn14}
\end{equation}
where 
\begin{equation}
g_{*}^{\rm eff}=g_{*}(T)\Bigg[1 + \frac{g_{*}(T_r)}{g_{*}(T)}\bigg(\frac{g_{*s}(T)}{g_{*s}(T_r)}\bigg)^{(4+n)/3} \bigg( \frac{T}{T_r}\bigg)^{n} \Bigg].
\label{eqn15}
\end{equation}
Using the Friedmann equation, we can write down the Hubble parameter as
\begin{equation}
H (T)=\sqrt{\frac{8 \pi G \rho(T)}{3}}=\frac{\sqrt{\rho(T)}}{\sqrt{3}M_{\rm Pl}}
\label{eqn16}
\end{equation}
which can now be rewritten using the effective relativistic degrees of freedom shown in \eqref{eqn15} as
\begin{equation}
H(T)= \frac{1}{\sqrt{3}M_{Pl}}\Bigg[ \frac{\pi^2}{30}g_{*}(T) T^4 \Bigg\{1 + \frac{g_{*}(T_r)}{g_{*}(T)}\bigg(\frac{g_{*s}(T)}{g_{*s}(T_r)}\bigg)^{(4+n)/3} \bigg( \frac{T}{T_r}\bigg)^{n}\Bigg\}  \Bigg]^{1/2}.
\end{equation}
In the above expression, $M_{\rm Pl}=(8\pi G)^{-1/2}=2.4 \times 10^{18}$ GeV is the reduced Planck mass. At $T \gg T_{r}$, assuming $g_* (T) = \bar{g}_* =$ constant for simplicity, the Hubble parameter can be approximated as 
\begin{equation}
H\approx\frac{\pi}{3}\frac{\bar{g_{*}}^{1/2}}{\sqrt{10}}\frac{T^2}{M_{\rm Pl}}\bigg(\frac{T}{T_r}\bigg)^{n/2}
\label{eqn17}
\end{equation}
Similarly, for entropy density, we can write
$$ s = \frac{\rho+p}{T} = (1+\omega) \frac{\rho}{T}$$
where $p = \omega \rho$ is the equation of state. Using the expressions for energy density from above discussion, the entropy density corresponding to $\phi$ field is
\begin{align}
s_{\phi} = (1+\omega) \rho_{\phi}(T_r) \bigg(\frac{g_{*s}(T)}{g_{*s}(T_r)}\bigg)^{(4+n)/3} \bigg( \frac{T}{T_r}\bigg)^{(4+n)} \frac{1}{T}
\end{align}
while for the usual radiation it is given by 
\begin{align}
s_{\rm rad} = \frac{2\pi^2}{45} g_{*s}\,T^3.
\end{align}
Therefore, the net effective entropy density is given by
\begin{align}
s & = s_{\rm rad} + s_{\phi} \nonumber \\
& = s_{\rm rad} \bigg [ 1+ \frac{3}{4}(1+\omega) \frac{g_* (T_r)}{g_{*s} (T)} \bigg(\frac{g_{*s}(T)}{g_{*s}(T_r)}\bigg)^{(4+n)/3} \bigg( \frac{T}{T_r}\bigg)^{n} \bigg].
\end{align}
Thus, we can write the effective entropy degrees of freedom as
\begin{equation}
g^{\rm eff}_{*s} (T) = g_{*s} (T) \bigg [ 1+ \frac{3}{4}(1+\omega) \frac{g_* (T_r)}{g_{*s} (T)} \bigg(\frac{g_{*s}(T)}{g_{*s}(T_r)}\bigg)^{(4+n)/3} \bigg( \frac{T}{T_r}\bigg)^{n} \bigg ] .
\label{g*seff}
\end{equation}
The equation of state parameter $\omega$ can be found for each value of $n$ by comparing the evolution of energy density 
\begin{equation}
\rho_{\phi} (T) = \rho_{\phi} (T_r) \left( \frac{a(T_r)}{a(T)} \right)^{3(1+\omega)}
\end{equation}
with the scaling mentioned earlier $\rho_{\phi} \propto a^{-(4+n)}, \; n>0$. For example, $n=1$ corresponds to $\omega = \frac{2}{3}$. Similarly $n=2,3,4$ correspond to $\omega =  1,  \frac{4}{3},  \frac{5}{3}$ respectively.

\begin{figure}[!h]
\centering
\begin{tabular}{cc}
\epsfig{file=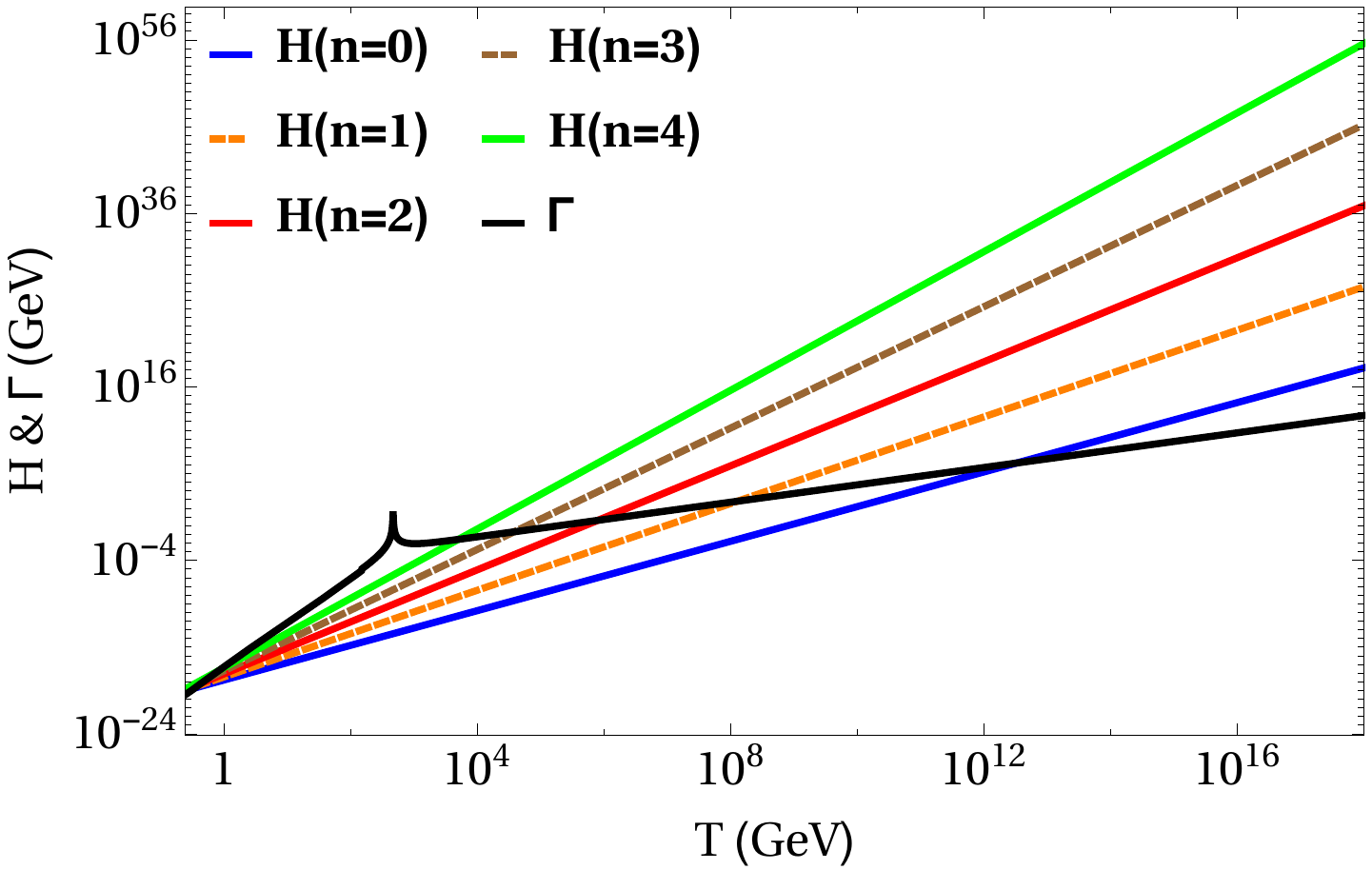,width=0.5\textwidth,clip=}
\epsfig{file=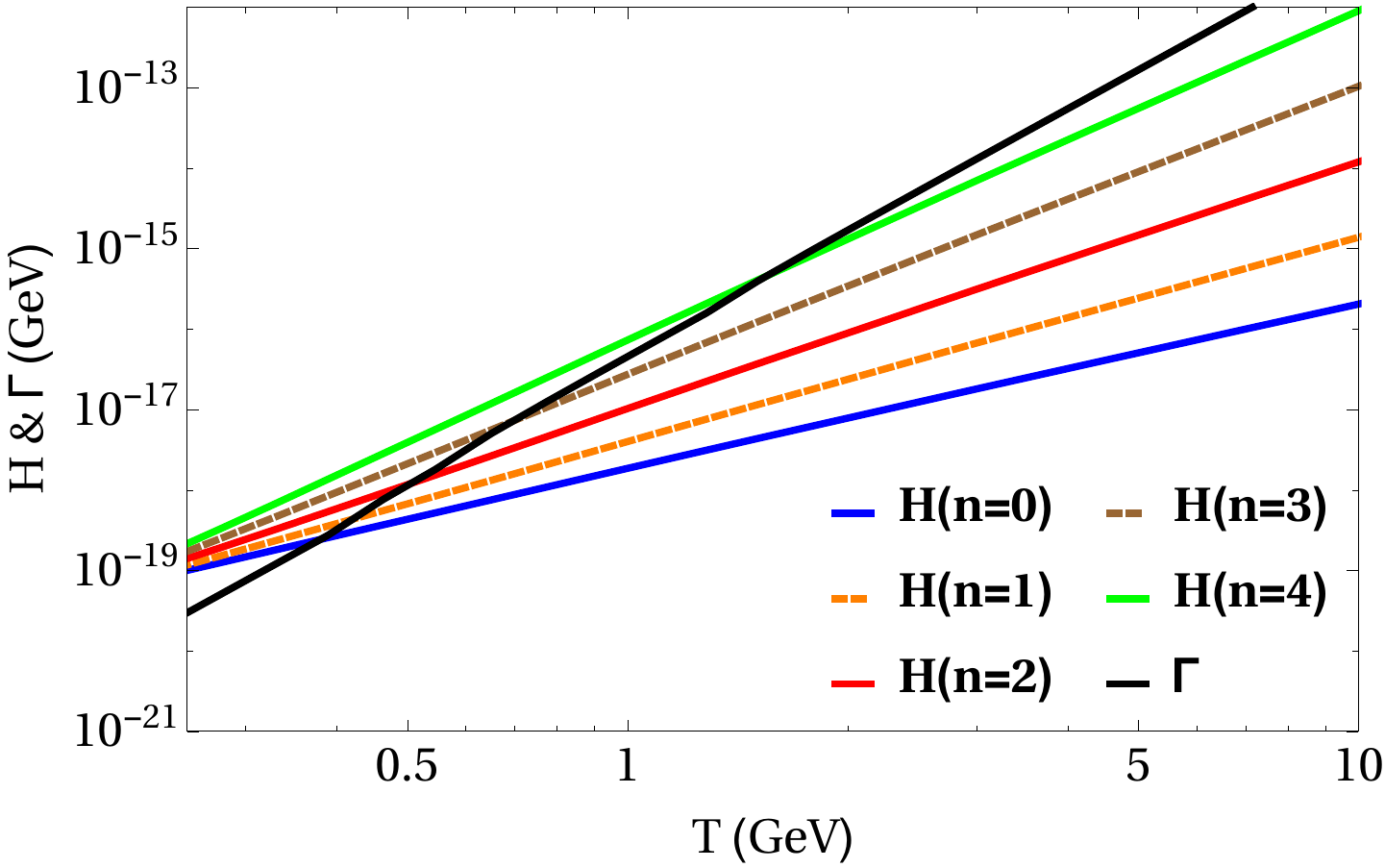,width=0.5\textwidth,clip=}

\end{tabular}
\caption{Left panel: Hubble parameter and interaction rate as a function of temperature for different values of n. We have kept fixed $m_{Z_{BL}}$=3 TeV, $g_{BL}$=0.2, and $T_{r}$=200 MeV. Right panel: Zooming in the low temperature regime of the same plot in left panel.}
\label{fig2}
\end{figure}

Now it is clear from the above equation \eqref{eqn17} that for $T \gg T_r$, Hubble parameter or the rate of expansion of the Universe is larger than what would have been for standard cosmology with $n=0$. As the Universe expands faster than the standard scenario, the decoupling of DM ($N_1$ in our model) will also happen at an earlier epoch. In figure \ref{fig2} we have shown the interaction rate of DM as well as Hubble parameter for different values of $n$. As can be seen from this figure, the DM interactions decouple from equilibrium plasma at earlier epochs for larger values of $n$. Another interesting feature is that the DM interactions were not always in equilibrium in the early Universe, it enters thermal equilibrium at some epochs and then departs at a later epoch. The duration between these two epochs is the shortest for larger values of $n$. From the behaviour of interaction rates and Hubble parameter in figure \ref{fig2}, it is clear that if we choose even larger values of $n \gg 4$ or different benchmark for $U(1)_{B-L}$ parameters, it may be possible that the DM interactions never attain thermal equilibrium in the early Universe. We however, do not discuss such a case in this work and leave it to future studies.

\section{keV dark matter in a fast expanding universe}
\label{sec:kevfast}
In section \ref{sec:kevmodel} we have mentioned the over-abundance problem of keV DM in this model where we have taken $m_{\rm DM}$ = 5 keV. We have also shown that to overcome this problem people have introduced the concept of entropy dilution in the early Universe which can be achieved by the out of equilibrium decay of some heavy particles after DM freezes out from the thermal bath. Here we propose a different possibility, namely the non-standard cosmological phase described in the previous section, to solve the over-abundance problem.

\begin{figure}[!!!!!!h]
\centering
\begin{tabular}{cc}
\epsfig{file=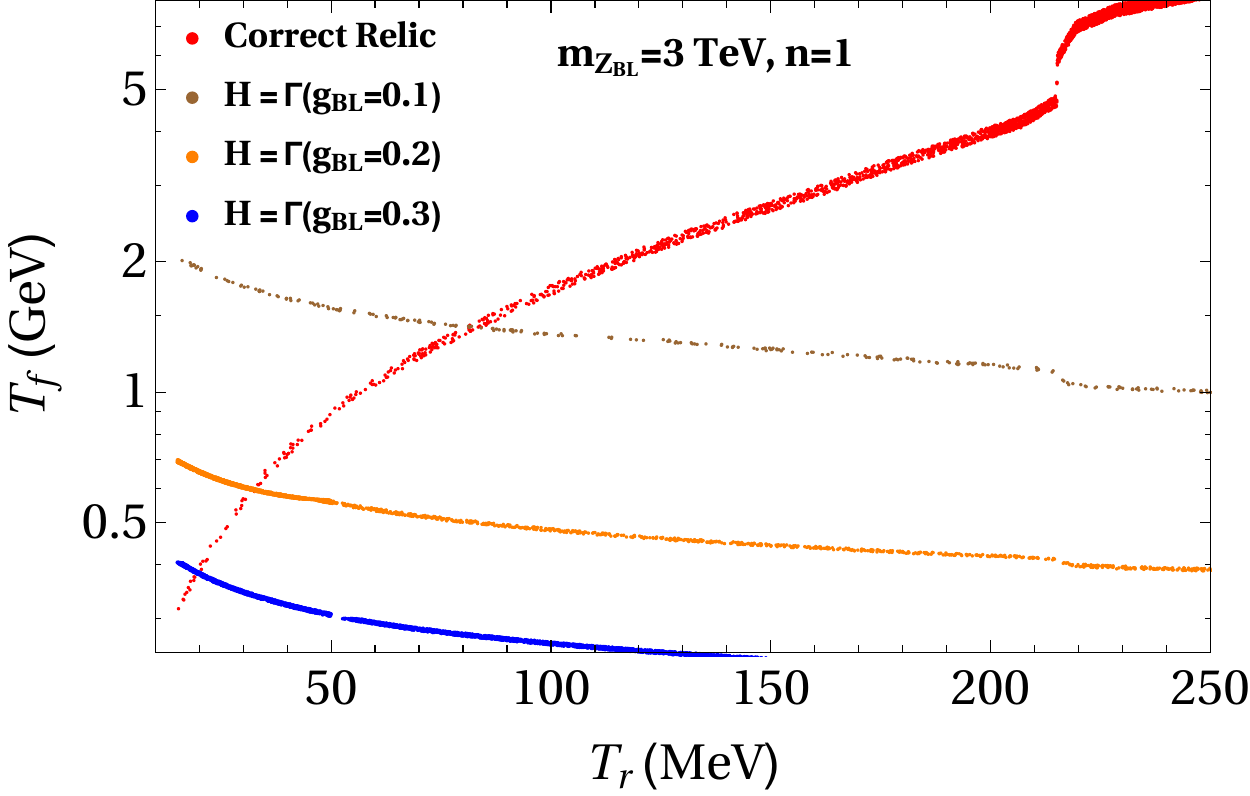,width=0.5\textwidth,clip=}
\epsfig{file=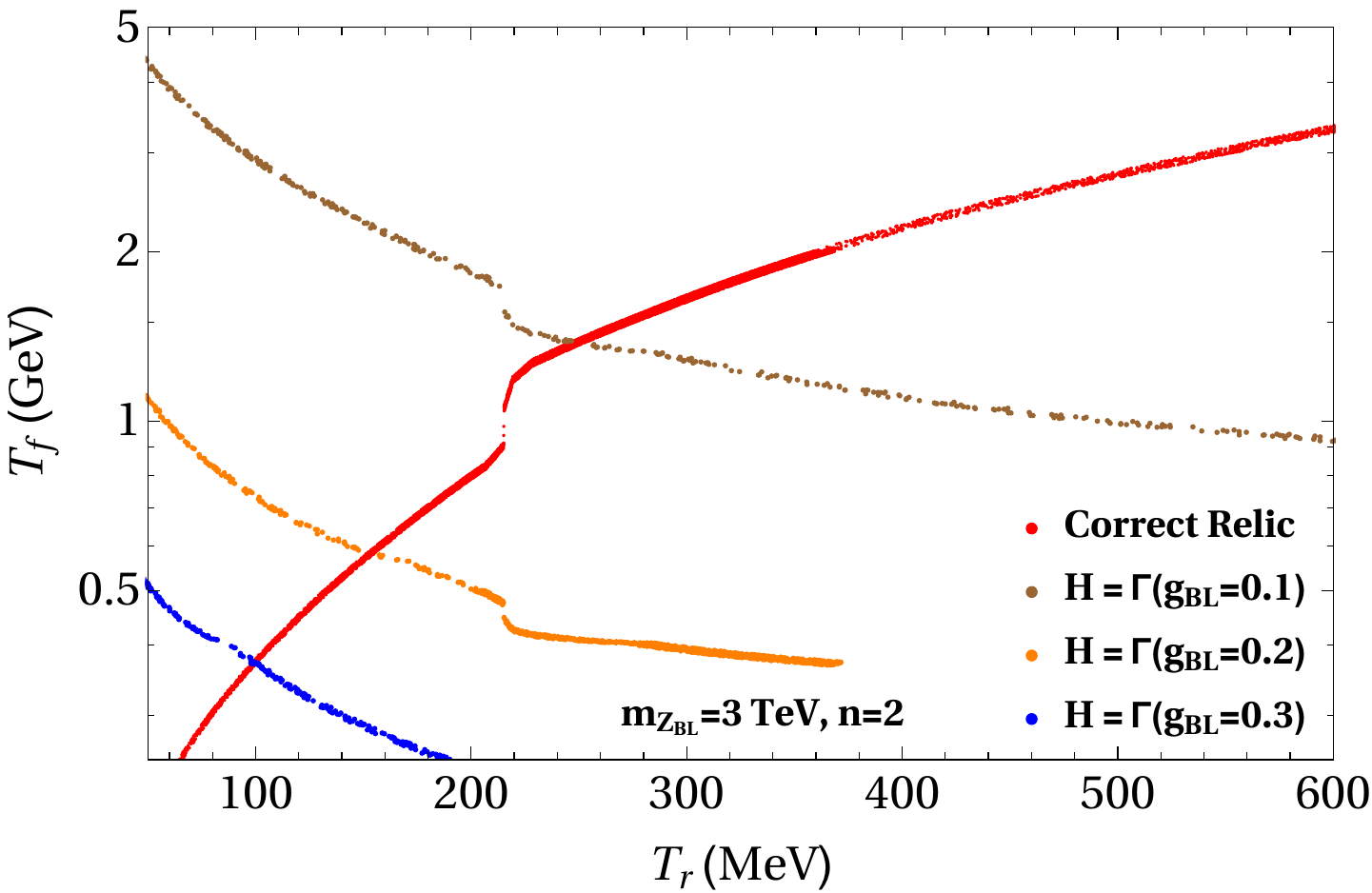,width=0.5\textwidth,clip=}
\end{tabular}
\caption{Allowed parameter space in $T$ vs $T_{r}$ plane, where the red points are giving the required values of $g_{\rm *s}^{\rm eff}$ to get the correct DM relic density. The brown, orange, and blue points represent the decoupling temperature $T_f$ of DM as a function of $T_r$ for benchmark values of $g_{BL}$=0.1, 0.2, 0.3 respectively. We have kept fixed $n=1,2$ and $m_{Z_{BL}}$=3 TeV.}
\label{fig3}
\end{figure}

\begin{figure}[!!!!!!h]
\centering
\begin{tabular}{cc}
\epsfig{file=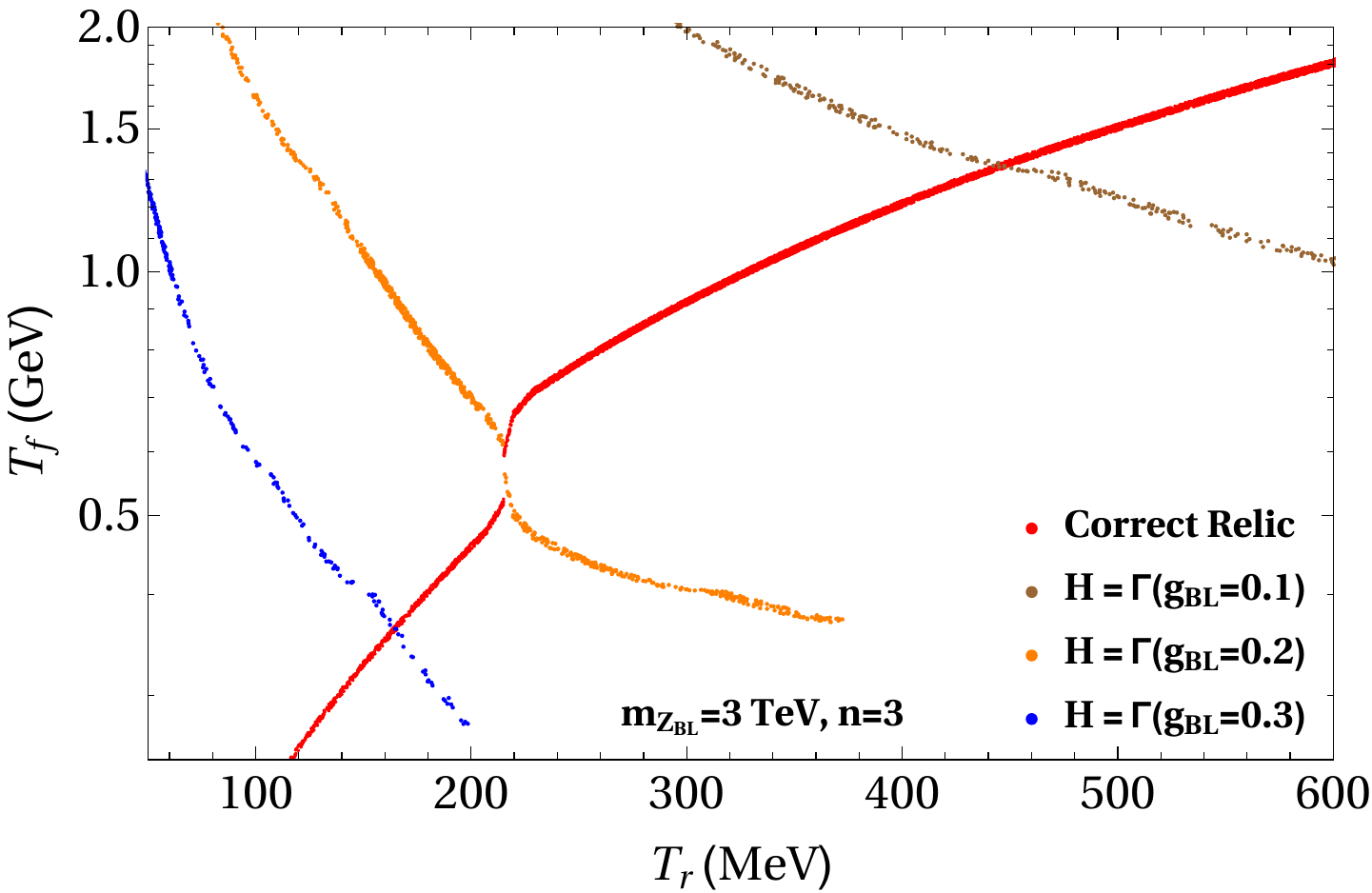,width=0.5\textwidth,clip=}
\epsfig{file=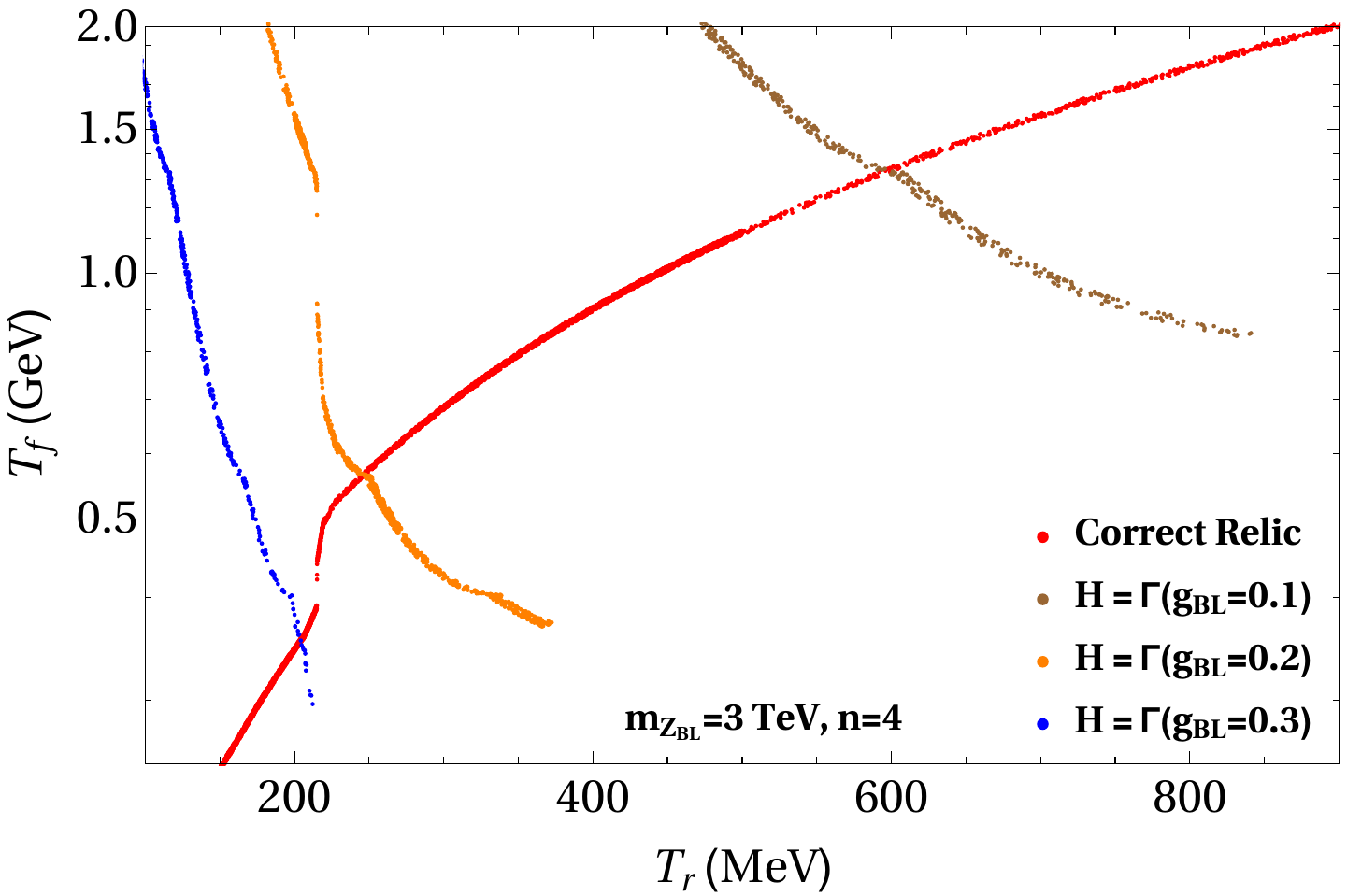,width=0.5\textwidth,clip=}
\end{tabular}
\caption{Allowed parameter space in $T$ vs $T_{r}$ plane, where the red points are giving the required values of $g_{\rm *s}^{\rm eff}$ to get the correct DM relic density. The brown, orange, and blue points represent the decoupling temperature $T_f$ of DM as a function of $T_r$ for benchmark values of $g_{BL}$=0.1, 0.2, 0.3 respectively. We have kept fixed n=3,4 and m$_{Z_{BL}}$=3 TeV.}
\label{fig3a}
\end{figure}

Let us assume that DM decouples from the thermal bath at some temperature $T>T_{r}$ when the expansion rate was faster than the radiation dominated phase of standard cosmology. Now, in the present scenario, there exists
a constraint on $T_r$ and $n$ from the number of relativistic degrees of freedom present during
BBN and this is given by \cite{DEramo:2017gpl},
\begin{eqnarray}
T_r \gtrsim (15.4)^{1/n}\,\,{\rm MeV}\,.
\end{eqnarray}
Therefore, one can consider any values of $n$ and $T_r$ obeying
the above relation as well as the condition $T_f>T_r$. For the
latter condition, we have to choose the $U(1)_{B-L}$ model parameters
appropriately. Now, recall from equation \eqref{eqn6}
$$\Omega_{\rm DM}h^2=76.4 \bigg[\frac{3g}{4}\frac{1}{g_{*s}(x_{f})}\bigg]\left( \frac{M_{N_1}}{\rm keV} \right)\,\,.$$
Now in the non-standard cosmological scenario $g_{*s}=g^{\rm eff}_{*s}$ is a function of $T$ and $T_{r}$ as given by equation \eqref{g*seff} and we can get the required relic density by adjusting the $T$ and $T_r$ so that the correct value of $g^{\rm eff}_{*s}(x_{f})$ is obtained. However, one also needs to make sure that the DM decouples at a temperature $T_f > T_r$ where the value of $g^{\rm eff}_{*s}(x_{f})$ takes the value which satisfies the requirement of producing the correct DM relic. According to the Planck data $\Omega_{\rm DM}h^2=0.1186\pm0.0020$ \cite{Ade:2015xua} and that requires $ g^{\rm eff}_{*s}$ around 4850 using the above expression for DM relic, if $M_{N_1} = 5$ keV. This can be achieved for some combinations of $T$ and $T_r$, provided the decoupling temperature of $N_1$ falls within such values of $T$. Now the required $g_{*s}^{\rm eff}$ is almost 500 times the usual $g_{*s} =10.75$ of standard cosmology after QCD phase transition as we had discussed in section \ref{sec:kevmodel}. In figure \ref{fig3}, \ref{fig3a} we have shown the allowed parameter space in $T$ vs $T_r$ plane where the red points represent those combinations of $T$ and $T_r$ which gives the $4700 \lessapprox g_{*s}^{\rm eff} \lessapprox 5000$. This range of $g_{*s}^{\rm eff}$ is chosen so as to get the desired DM relic within the error bars mentioned in \eqref{dm_relic}.

On top of that we have to make sure that dark matter ($N_1$) also decouples from the thermal
bath at the same temperature and that can be confirmed by using the condition 
\begin{equation}
\Gamma (T) \approx H (T)
\label{eqn18}
\end{equation}
where as mentioned earlier in equation \eqref{gamma}, $\Gamma =n_{\rm DM} \langle \sigma {\rm v} \rangle$
is the interaction rate and $\langle \sigma {\rm v} \rangle$ is the thermally averaged
DM annihilation cross section which depends on the gauge coupling constant g$_{BL}$,
mass of the gauge boson m$_{Z_{BL}}$ (see equation \eqref{sigmav}). Moreover,
$n_{\rm DM}$ being the number density of our DM candidate $N_1$. 
By using equations \eqref{gamma}, \eqref{eqn16} and \eqref{eqn18}
one can write 
\begin{equation}
T^3 <\sigma v>\approx \frac{\sqrt{\rho(T)}}{\sqrt{3}M_{Pl}}
\label{eqn19}
\end{equation}
In figure \ref{fig3} we have shown the combinations of $T$ and $T_r$ which satisfy equation \eqref{eqn19} for different benchmark values of $g_{BL}$ (0.1, 0.2, 0.3) by keeping $m_{Z_{BL}}=3$ TeV and $n=1, 2$. The same is repeated for $n=3,4$ in figure \ref{fig3a}. As can be seen from these two plots, there are very small overlapping regions which satisfy both the relic density as well as will give the correct decoupling temperature. Another important point to note here is that both these plots contain the signature of the QCD phase transition below which the standard $g_{*s}$ suddenly drops and that can be understood from the sudden drop of decoupling temperature T$_f$ (red points) near few hundred MeV in figure \ref{fig3} and figure \ref{fig3a}. From equation \eqref{g*seff} it is clear that for $n>0$, $g_{*s}^{\rm eff}$ decreases with increase in $g_{*s} (T_r)$. As $T_r$ is decreased below around 200 MeV, there is a sudden drop in $g_{*s} (T_r)$ and hence sudden increase in $g_{*s}^{\rm eff}$. Therefore, in order to have the required value of $g_{*s}^{\rm eff} \in (4700-5000)$ at the decoupling temperature, there is a need for $T_f$ to drop to a lower value, as can be seen from figure \ref{fig3} and figure \ref{fig3a}. It is also clear that as we increase the coupling constant $g_{BL}$ for a fixed value of $m_{Z_{BL}}$, DM interacts more and stays in equilibrium for a longer time leading to a lower decoupling temperature. On the other hand, for smaller gauge couplings the decoupling occurs at much earlier epoch. For example, in figure \ref{fig3a}, DM for $g_{BL}=0.1$ decouples earlier compared to the cases for higher values of $g_{BL}$. Also, in both these plots, we can see that the correct DM relic is obtained for those values of $T_r$ which are much above the BBN temperature and hence are not going to be in conflict with successful BBN predictions. 

\section{Observational Prospects}
\label{sec:det}
As mentioned earlier, such a keV neutrino DM having tiny mixing with SM neutrinos by virtue of Yukawa couplings, can decay into a photon and light neutrino at radiative level with $W$ boson and charged leptons of the SM in loop. The corresponding decay width is given by \cite{Pal:1981rm}
\begin{equation}
\Gamma (N_1 \rightarrow \nu \gamma) \approx 1.38 \times 10^{-29} \; \text{s}^{-1} \left ( \frac{\sin^2{2\theta}}{1\times 10^{-7}} \right) \left ( \frac{M_{N_1}}{1 \; \text{keV}} \right)^5
\end{equation}
where $\theta$ denotes the mixing between $N_1$ and $\nu$. From the observation of the 3.55 keV line, which can arise from the decay of a 7.1 keV sterile neutrino DM, the mixing angle which is in agreement with the observed flux is $\sin^2{2 \theta} \approx 7 \times 10^{-11}$ \cite{Bulbul:2014sua}. Such a mixing angle can be realised in our model by suitable adjustment of Yukawa couplings $y^{\nu}$. It should be noted that apart from this minimal framework explaining the 3.55 keV line, there exists other scenarios as well. Different possible keV DM scenarios that can give rise to such an X-ray line have been studied, see for example \cite{Arcadi:2014dca}. One can also generate such a signal in typical WIMP DM models if there are two quasi-degenerate DM candidates having mass splitting of 3.55 keV, allowing the heavier one to decay into the lighter one and a photon. Such possibilities were discussed in \cite{Borah:2015rla, Borah:2016ees} among others. The alternative possibility of keV dark matter annihilation into monochromatic photons was also discussed very recently by the authors of \cite{Brdar:2017wgy}. Although the analysis of the preliminary data collected by the Hitomi satellite (before its unfortunate crash) do not confirm such a monochromatic line \cite{Aharonian:2016gzq}, one still needs to wait for a more sensitive observation with future experiments to have a final word on it.

Another observational aspect of such keV DM is the astrophysical structure formation rates which is determined by the free streaming length. In particular, depending upon the DM free streaming lengths, the Lyman-$\alpha$ forest observations may get affected. The Lyman-$\alpha$ forest, produced by filaments of neutral hydrogen in the intergalactic medium along the line of sight to a distant quasar can depend upon the coldness of the DM which is present along with the baryonic fluid. For example, a hot DM component will broaden the lines while a cold component will make them more compact. For warm dark matter scenario like ours, where the momentum distribution of DM particles is proportional to a thermal spectrum, the free streaming length of DM can be estimated as \cite{Bond:1980ha}
\begin{equation}
\lambda_{\rm FS} \approx 1 {\rm MPc} \frac{\rm keV}{m_{\rm DM}} \frac{ \langle p_{\rm DM} \rangle}{\langle p_{\nu} \rangle }
\end{equation}
where $\langle p_{\rm DM} \rangle$ is the average momentum of DM particles and $\langle p_{\nu} \rangle \sim 1$ keV is the comoving momentum of light neutrinos at the epoch when thermally produced DM ($N_1$ with keV mass in this case) particles become non-relativistic $(T \leq M_{\rm DM} \equiv M_{N_1})$. For non-thermal origin of DM, such a simple formula for free streaming length is not valid and details of such cases can be found in the recent review \cite{Adhikari:2016bei} as well as the references therein. As can be seen from the above formula, for thermally produced sterile neutrino DM with mass in the few keV range, the free streaming length can be within a Mpc, as required by the Lyman-$\alpha$ forest data.

\section{Conclusions}
\label{sec:conc}
We have studied the possibility of having a keV sterile neutrino dark matter in the minimal $U(1)_{B-L}$ gauge extension of the standard model. The anomaly cancellation requirements naturally allow the existence of three right handed neutrinos having $U(1)_{B-L}$ charge $-1$ each in the model, allowing the realisation of type I seesaw mechanism for generating light neutrino masses. Due to the absence of any remnant symmetry protecting the stability of a dark matter candidate in the minimal model, the only way to have a dark matter candidate is to forbid its decay kinematically by keeping its mass in the keV regime. The lightest right handed neutrino can do such a job if it has tiny mixing with the light neutrinos that gives rise to its cosmologically long lifetime. Such tiny mixing of a keV right handed neutrino effectively decouples it from the light neutrino mass generation mechanism, leaving the lightest neutrino mass vanishingly small.

Due to the gauge interactions of the right handed neutrinos, the keV dark matter can be thermally produced in the early Universe and typically gets overproduced if the $U(1)_{B-L}$ gauge boson mass lies in the TeV corner with sizeable gauge coupling. Instead of adopting the usual approach of late entropy dilution in order to bring the keV neutrino abundance under control, we consider a non-standard cosmological phase dominated by a scalar field $\phi$ whose energy density scales faster than radiation $\rho_{\phi} \propto a^{-(4+n)}, n>0$. Such a phase which ends prior to the era of the big bang nucleosynthesis, gives rise to a different temperature dependence of relativistic degrees of freedom compared to the standard cosmology. We constrain the model parameters as well as the scaling power $n$ considering positive integral values for which the keV neutrino can be thermally produced in the early Universe but at the same time does not lead to overproduction at late times. Such a scenario can leave much more parameter space of the model allowed as it does not constrain the masses and Yukawa couplings of the heavier right handed neutrinos like in the entropy dilution approach. As pointed out in one such earlier work \cite{Nemevsek:2012cd}, the correct late entropy dilution and the requirement of not producing the keV neutrino in the late decays of heavier neutrinos (which inject too much energy to the DM at late epochs, leading to the erasure of structures at scales above a Mpc) severely constrain the model parameters, which may not be consistent with the generation of light neutrino masses in the minimal $U(1)_{B-L}$ model like the one we discuss here. The model can therefore survive the phenomenological tests of correct DM abundance and light neutrinos masses if such non-standard cosmological phase is taken into account. As long as the heavier right handed neutrinos decay before the keV neutrino decouples, which can be realised naturally for generic Yukawa coupling and TeV scale heavy neutrino masses, the DM phenomenology is not going to depend upon the spectrum of heavier neutrinos.

We also comment upon the possibility of generating the unidentified 3.55 keV X-ray line claimed to be present in the XMM-Newton data in this model from radiative decay of a 7.1 keV dark matter into a photon and a light neutrino. Also, the free streaming length of such thermally generated keV dark matter can remain in the warm dark matter regime which has the potential to resolve the small scale structure issues that exist in the cold dark matter paradigm.

\acknowledgments
The authors would like to thank the members of HEP Journal Club (JC) of IIT Guwahati for useful discussions during a JC talk where this idea was conceived. DB acknowledges the support from IIT Guwahati start-up grant (reference number: xPHYSUGIITG01152xxDB001) and Associateship Programme of IUCAA, Pune. 

\bibliographystyle{apsrev}

\end{document}